\documentclass[aps,twocolumn]{revtex4}
\usepackage{epsfig}
\usepackage{amsmath}
\usepackage{epsfig}

\begin{document}
\title
{\bf An update on coherent scattering from complex non-PT-symmetric Scarf II potential with  new analytic forms}   
\author{ Sachin Kumar$^1$ and Zafar Ahmed$^{2,3}$}
\email{1: sachinv@barc.gov.in, 2:zahmed@barc.gov.in, zai-alpha@hotmail.com}
\affiliation{$~^1$Theoretical Physics Section, $~^2$Nuclear Physics Division, Bhabha Atomic Research Centre,Mumbai 400 085, India \\
	$~^3$Homi Bhabha National Institute, Anushaktinagar, Mumbai 400094, India}
\date{\today}
\begin{abstract}
The versatile and exactly solvable Scarf II has been predicting, confirming and demonstrating interesting phenomena in complex PT-symmetric sector, most impressively. However, for the non-PT-symmetric sector it has gone underutilized. Here, we present most simple analytic forms for the scattering coefficients $(T(k),R(k),|\det S(k)|)$. On  one hand, these forms  demonstrate earlier effects and confirm the recent ones. On the other hand they  make new predictions - all  simply and analytically. We show the possibilities of both self-dual and non-self-dual spectral singularities (NSDSS) in  two non-PT sectors (potentials). The former one is not accompanied by time-reversed coherent perfect absorption (CPA)  and gives rise to the parametrically controlled splitting of SS in to a finite number of complex conjugate pairs of eigenvalues (CCPEs). The latter ones (NSDSS) behave just oppositely: CPA but no splitting of SS. We demonstrate a one-sided reflectionlessness without invisibility. Most importantly, we bring out a surprising co-existence of  both real discrete spectrum and a single SS in a fixed potential. Nevertheless, the complex Scarf II is not known to be pseudo-Hermitian ($\eta^{-1} H\eta=H^\dagger$) under a metric of the type $\eta(x)$, so far.
\end{abstract}
\maketitle
\section{Introduction}
An  exactly solvable Hermitian potential called  Scarf II  proposed by  Gendenshtein [1] is written as 
\begin{multline}
V_{S_1}(x)= P ~ \mbox{sech}^2x\ +  Q ~\mbox{sech}x \tanh x, \\
 P= (B^2-A^2-A),~ Q= B(2A+1).
\end{multline}
After the discovery [2] of real spectrum of complex PT-symmetric Hamiltonians, this potential became the most interesting  exactly solvable complexified model  $(B=i B')$ [3] of the new quantum mechanics. In another complexification: $P=-V_1, Q=iV_2, V_1,V_2 \in {\cal R}, V_1>0$, it additionally had the distinction of having both  the  exact and broken regimes of PT-symmetry according to whether $|V_2| \le V_1+1/4$ [4].
Remarkably, the exactly solvable Scarf II (1) is a localized potential which is a very  useful scattering potential. In 1988, Khare and Sukhatme have obtained beautiful expressions of its transmission $t(k)$ and reflection $r(k)$ amplitudes in terms of Gamma functions with complex argument, where the parameters are real and $V(x)$ is Hermitian [5]. For about twenty years or so the complexified Scarf II has been re-enforced for the novel features of a new kind of scattering from it.
\begin{figure}
\includegraphics[width=4.2 cm,height=4.5 cm]{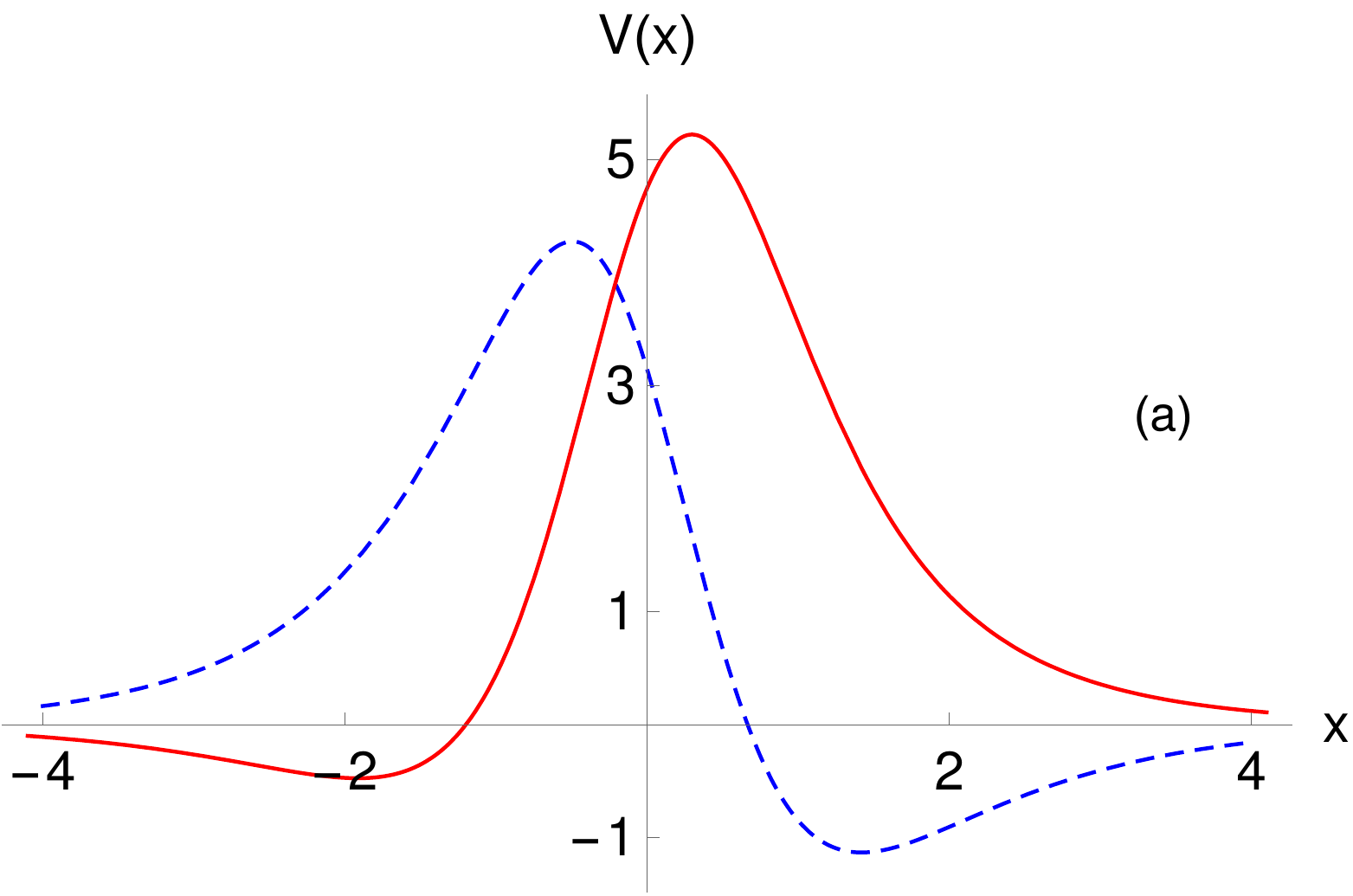}
\includegraphics[width=4.2 cm,height=4.5 cm]{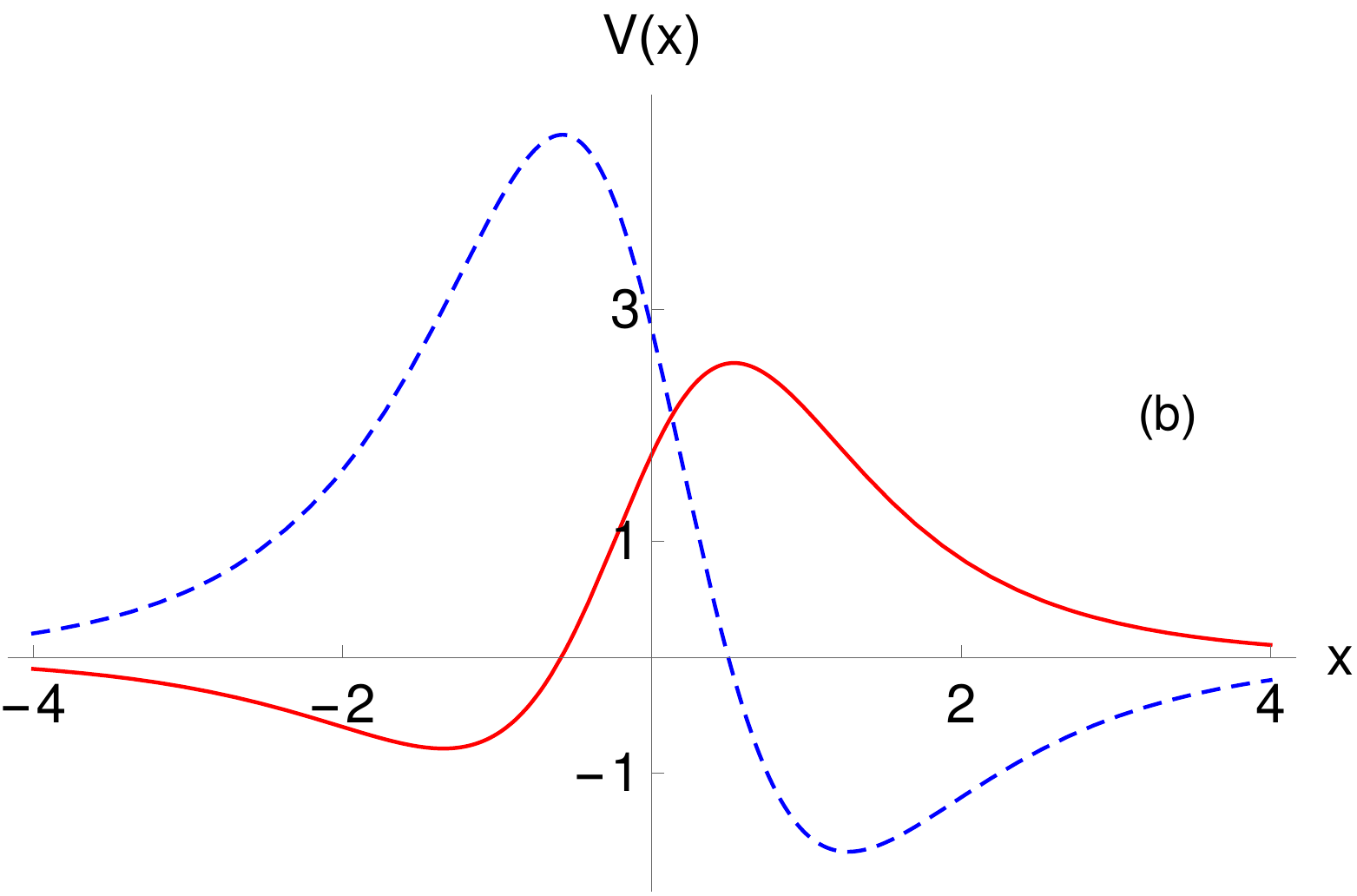}
\caption{ (a): Scarf II $V(x)$ (1) for  $A=-ic, B=d+i/2, c=\sqrt{2}, d=\sqrt{3}$ (see section II). (b): Scarf II $V(x)$ (1) for $ A=1-ic, B=c-i/2, c=\sqrt{2}$ (see section III). Solid (red) line is for real part and dashed (blue) line is for imaginary part. }
\end{figure}

In 2001, it was stated and proved [6] that if a scattering potential is complex and asymmetric the reflection probability show left/right handedness, later
this handedness became an essential feature of complex PT-symmetric potentials and the PT-sector of  (1) presented  a very interesting analytic demonstration [7] of the now called non-reciprocal reflection. This opened up the scope of studying coherent scattering where
two identical beams of particles are sent from left and right at the non-Hermitian potential (optical medium). By PT-sector we mean the parametric regime of (1) wherein $(V(-x, -i)=V(x,i))$.

Two crucially important concepts/tools have been developed, one is the determinant of two port S-matrix [8]
\begin{equation}
|\det S(k)|=|t^2(k)-r_{left}(k)~ r_{right}(k)|,~ E=k^2
\end{equation}
and the other one is the  spectral singularity(SS) [8], wherein at the incident energy $E=E_*=k_*^2$ all three scattering coefficient $R_{left}(k), R_{right}(k)$ and $T(k)$  become infinite. Complex PT-symmetric Scarf II potential turned out to an exactly solvable model [9,10] of SS explicitly.
The most interesting physical  interpretation of SS  is given by the fact that the SS at $E=E_*$
in $T(k)$ leads to  a zero in the determinant of  the time reversed S-matrix (2) as $|\det S(-k_*)|=0$. This phenomena is called coherent perfect absorption (CPA) which in turn leads to time-reversed lasers [11]. When the medium is PT-symmetric and this symmetry is broken another phenomena called CPA with lasing has been observed [12]. If an SS occurs in both $T(k)$ and  $T(-k)$ it is called self-dual, this was first discovered [8] in the PT-sector, later it has been proposed to occur in the non-PT sector as well [13]. Scarf II potential (1) had  already been found [13] to have both  SDSS and NSDSS in PT and non-PT sectors, respectively.
In the PT-sector, if the potential is set for an SS then a slight increase in the strength of the  imaginary part of $V(x)$ gives rise to the splitting [15] of SS into a complex conjugate pair of eigenvalues (CCPEs). It is worth noting that discovery  of this phase transition [15] owes it to an interesting hint coming from Scarf II (1) [10]. Further increase of this strength causes more and more number of CCPEs. In PT-sector real-discrete spectrum and SS are found to be mutually exclusive, SS and CCPEs occur strictly in the domain of broken PT-symmetry, then SS is conjectured to be single which sets an upper bound to the finite number of CCPEs: $E_* >( \approx) \Re(E_l)$ [16], where ${\cal E}_l$ is that last of CCPEs when arranged w.r.t. their real parts. Recently, the splitting of SS has been found to occur also in a non-PT sector [17].  Earlier, two NSDSS [14] were found in a non-PT Scarf II (1). A construction [17] of numerically solved non-PT potentials having two parametrically controlled SSs (one SDSS and one NSDSS) has been proposed. Designing of Lasing and perfectly absorbing potentials is being searched [18]  meticulously both theoretically and experimentally.
\begin{widetext}
The beautiful transmission amplitude of Scarf II (1) found by Khare and Sukhatme [5] is given as
\begin{equation}
t_{A,B}(k)=\frac{\Gamma[-A-ik] \Gamma[1+A-ik] \Gamma[\frac{1}{2}+iB-ik] \Gamma[\frac{1}{2}-iB-ik]}{\Gamma[-ik] \Gamma[1-ik] \Gamma^2[\frac{1}{2}-ik]},
\end{equation}
It  can readily be checked that $t_l=t_{A,B}$ and $t_r=t_{A,-B}$ are equal representing the reciprocity of transmission can be noted. The reflection amplitude is given as
\begin{equation}
r_{A,B}(k)=t_{A,B}(k) f_{A,B}(k),
\end{equation}
\begin{equation}
f_{A,B}(k)=\left [\frac{\cos \pi A \sinh \pi B}{\cosh \pi k}+ i \frac{\sin \pi A \cosh \pi B}{\sinh \pi k} \right ],
\end{equation}
\end{widetext}
so in view of the non-reciprocity of reflection [6,7], we find that  $r_{l}=r_{A,B}, r_{r}=r_{A,-B}$.
Let us also introduce $F_l=f_{A,B}$ and $F_r=f_{A,-B}$ to be used in the sequel.
When  $|F_l|=|F_r|$  the reflection is reciprocal: $R_l=R_r$, this happens when $A$ and $B$ are real ($V(x)$ is Hermitian) or when they are purely imaginary. The second possibility is surprising as $V(x)$(1) would be both non-Hermitian and spatially non-symmetric (see Eq. (23) in section IV, below). Otherwise, in general the reflection for non-Hermitian and non-symmetric potential is non-reciprocal $R_l \ne R_r$ as stated and proved in [6].
\begin{figure}
	\includegraphics[width=7.5 cm,height=5 cm]{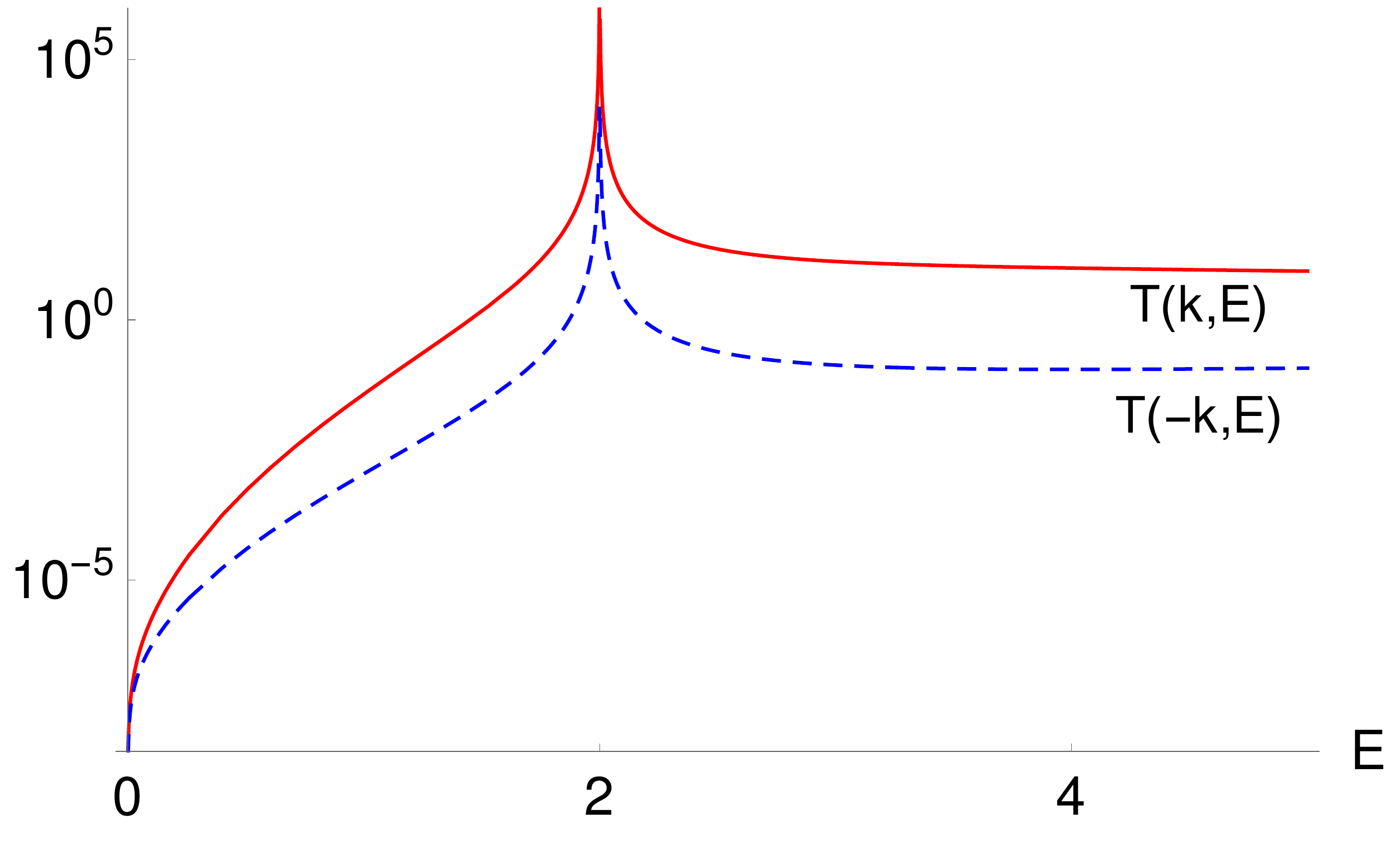}
	\caption{Depiction of scenario for the self-dual spectral singularity in non PT-sector
		($A=1-ic, B=c-i/2, c=\sqrt{2}$). Both $T(k,E)$ and $T(-k,E)$ are un-equal but infinite at the same point $k=c$ (See Eq. (13)). In PT-sector one always gets $T(k,E)=T(-k,E)$ [14,20].}
\end{figure}
When either $A$ or $B=iB'$ are purely imaginary $f_{A,B}$ can become zero for instance at
\begin{multline}
k_z=\pm \pi^{-1} \tanh^{-1} \left(\tan \pi A \coth B'\right), \\ \quad \mbox{if} \quad -1 \le \tan \pi A \coth B' \le 1.
\end{multline}
This unidirectional reflectionlessness  turns out to be unidirectional invisibility [18] provided $T(k_z)=1$.

We would like to explain as to how we could successfully eliminate $\Gamma(z)$ functions with complex arguments while calculating simple analytic forms of $T(k)=|t(k)|^2=t(k)\bar t(k)$ in this paper. We do this by using $\Gamma(z) \Gamma(1-z)=\pi \csc(\pi z)$. For instance $|\Gamma[-A-ik] ~ \Gamma[1+A-ik]|^2=\Gamma[-A-ik] ~ \Gamma[-A+ik] ~  \Gamma[1+A-ik] ~ \Gamma[1+A+ik]=\Gamma[-z_1]~\Gamma[- z_2] ~\Gamma[1+z_2] ~ \Gamma[1+z_1]$, combining   first term with fourth and second with third, we eliminate Gamma functions. Here $z_1=A+ik$ and $z_2=A-ik$. This method succeeds in the parametrizations: ${\bf P_1, P_2, P_3}$ and ${\bf P_4}$ discussed below where $A$ and $B$ are  parametrized specially. For other cases it may not be possible to get simple analytic forms for $T(k)$

In this paper, we present new simple analytic forms of $R(k),T(k), |\det(S(k)|$ in various non-PT sectors of Scarf II by  various complex parameterizations of $A$ and $B$ in (1). Earlier, this has been done for PT-sectors of (1) [20]. In this paper we reveal the contrasting features of SDSS and NSDSS. The former can be made to split by perturbing the potential but it does not give rise to CPA. We show the NSDSS that occurs in non-PT sector behave just oppositely. We show one sided reflectionlessness devoid of invisibility [19]. Most importantly, we show a coexistence of real discrete spectrum and the NSDSS in a fixed non-PT complex Scarf II (1). Nevertheless, this potential (1) in non-PT-sectors is not known to be  pseudo-Hermitian [22]: $\eta^{-1} H \eta= H^\dagger$, so far, under a local metric: $\eta(x)$.

In the following in sections II-V, we study and present 4 different parametric parameterization $({\bf P_1-P_4})$ of non-PT-sector of complex Scarf II (1), to bring out various features of coherent scattering simply and analytically. In sections VI and VII, we present the discussions and conclusions.
\begin{widetext} 
\section{\bf One/two non-self dual spectral singularity (NSDSS), no coherent perfect absorption (CPA),  one reflectivity zero (RZ)}
${\bf P_1:} A=-ic, B= d+ i/2$, from Eq. (1), the potential becomes 
\begin{equation}
V(x)=(c^2+d^2+i(c+d)-1/4)~ \mbox{sech}^2 x +(c+d-2icd+i/2)~\mbox{sech}x~\mbox{tanh}x, c,d \in R.
\end{equation}
By eliminating $\Gamma(z)$ functions using the identities like $\Gamma(z) \Gamma(1-z)= \pi z \mbox{cosec} \pi z$, trigonometric and hyperbolic identities in  Eq. (3), we get
\begin{equation}
T(k)=\left|\frac{(k+c)(k+d)}{(k-c)(k-d)}~\frac{(\sinh^2 \pi k \cosh^2 \pi k)}{(\cosh^2 \pi k-\cosh^2 \pi c)(\cosh^2 \pi k-\cosh^2 \pi d)}\right|.
\end{equation}
From Eqs. (4-5), we can write
\begin{equation}
F_{l,r}(k)=i \left [ \pm \frac{\cosh \pi c \cosh \pi d }{\cosh \pi k}
+ \frac{\sinh \pi c \sinh \pi d}{\sinh \pi k} \right]
\end{equation}
Clearly, two  spectral singularities exist at $k=k_{*1}=c$ and $k=k_{*2}=d.$
Next, there exists a single reflectivity zero as
\begin{equation}
k_z= \pi^{-1} \tanh^{-1} [\tanh \pi c \tanh \pi d].
\end{equation}
\end{widetext}
\vspace*{-0.5cm}
When $c$ is large such that $\tanh \pi c \approx 1$,
the $k_z \approx d$ and if $d$ is large then $k_z \approx c$. In these
cases the SS will appear to coincide with the single reflectivity zero. 
In the other cases $k_{*1},k_{*2}$ and $k_z$ will be distinct. The modulus of the determinant of two-port S-matrix (2)
after interesting simplifications becomes
\begin{equation}
|\det S(k)|=\left |\frac{(k+c)(k+d)}{(k-c)(k-d)}\right|,
\end{equation}
which becomes $\infty$ for $k=c,d$ and in time reversed setting  $T(-k)$ at $k=c, d$ it attains finite.So these two are NSDSS [13].

If $c>0$ ans $d<0$, From Eqs. (9,11) , it clearly turns out that there will be one SS in $T(E)$ at $k=k_*=c$ and one SS in the time reversed case: ($T(-k)$) at $k=k_*=d$, once again these are  NSDSS. Next,  when $c=d$, Eq. (8) shows that there will be only one SS which will again  be NSDSS occurring  at $k=k_*=c$  in  $T(k)$ but not in $T(-k)$. At $k=c$, the former is infinite but the latter attains a  finite limit.  The case of $c=-d$, makes the potential (1) PT-symmetric [19] and from Eqs. (9,11) one can  readily get the  SDSS at $k=\pm c =k_*$ and the  CPA with lasing characteristically [11]   as $|\det S(\pm c)|=\frac{0}{0}$, but $\lim_{k \rightarrow \pm c} |\det S(k)|=1$.
With this we can also rule out CPA with lasing in a non-PT sector.
\vspace*{-0.4 cm}
\section{\bf \hspace{-0.2 cm }one self-dual SS, no CPA, and one RZ}
${\bf P_2:} A=1-ic, B=c-i/2,$  the potential (1) becomes
\begin{multline}
\hspace{-0.8 cm}V(x){=}(2c^2{+}2ic{-}\frac{9}{4}){\mbox{sech}^2 x}{+}(2c{-}2ic^2{-}\frac{3i}{2}){\mbox{sech}x ~\mbox{tanh}x}, c  \in R.
\end{multline}
Further, we can convert (3-5) to  simple analytic forms as
\begin{equation}
T(k)=\left|\frac{(1+(k+c)^2)}{(1+(k-c)^2)}\frac{(\sinh^2 \pi k \cosh^2 \pi k}{(\cosh^2 \pi k-\cosh^2 \pi c)^2} \right|.
\end{equation}
Notice that $T(k) \ne T(-k)$, but they singular at the same point $E=c^2$. See Fig. 3.
In PT- sector one gets $T(k)=T(-k)$ [13,19]. 
\begin{equation}
F_{l,r}(k)=i \left[\pm \frac{\cosh^2\pi c}{\cosh \pi k} +\frac{\sinh^2\pi c}{\sinh \pi k} \right]
\end{equation}
and (2) as
\vspace{-0.5 cm}
\begin{equation}
|\det S(k)|=\frac{1+(k+c)^2}{1+(k-c)^2},
\end{equation}
which rules out both CPA with [9] or without [11] lasing.

\vspace{-0.7 cm}
\begin{equation} 
k_z=\pm \frac{1}{\pi}\tanh^{-1}[\tanh^2(\pi c)]
\end{equation}
\vspace*{-1.1 cm}
\section {\bf Splitting of SDSS into complex conjugate pairs of eigenvalues (CCPEs), no CPA, one RZ}

${\bf P3:} A=q+1/2-ic, B=c-iq, q\ge 1/2, c>0$
For this phenomenon, we have  introduced a new parameter $p$  to control the splitting of the self dual spectral singularity.  In this regard, the previous parametrizations: 
${\bf P_1, P_2}$ can be seen to be the cases of $q=-1/2,1/2$, respectively.
For ${\bf P4}$, the potential becomes
\begin{multline}
V(x)=(2c^2-2q^2+2ic-2q-3/4)~ \mbox{sech}^2 x \\ + [2c-2ic^2-2iq-2iq^2)~ \mbox{sech}x \tanh x
\end{multline}
Complex PT-symmetric potentials have been discussed to have three types [16] of discrete spectrum which come as complex k-poles of $t(k)$. the physical poles are of the types: $k=\pm k_{x}, ik_y, \pm k_x+ik_y ~ (k_y >0)$ for SS, bound-states and complex conjugate eigenvalues, respectively. Other poles are un-physical. The recent  phenomena of splitting of SS in to CCPEs in non-PT-sector gives the opportunity to discuss the presence of a finite number of CCPEs in non-PT sector of (i) with ${\bf P_3}.$ In this regard,  the four parameter dependent Gamma functions  in the numerator of (3) are interesting.We use  $\Gamma{-n}=\infty, n=0,1,2,..$, the the first and last one gives finite number of poles as $k_m=c +(q+1/2-m)i, k_n=-c+(q-1/2-n)i, m,n=0,1,2..$ $m \le q+1/2$ and $n \le q-1/2$, such that $k_y \ge 0$. These pairs of poles can be synchronized as.
%\vspace*{-.5 cm}
\begin{multline}
k_n=\pm c+(q-1/2-n), n=0,1,2,..\le [q-1/2] ~\mbox{and}~ \\ k^{\uparrow}= c+(q+1/2)i, k_{0} (q=1/2)=c=k_*
\end{multline}
%\vspace{-.5 cm}
With $k^{\uparrow}$  being single (unpaired). The middle two Gamma functions in (3) give rise to infinitely many pairs of un-physical poles ($k_y<0$), unlike the poles in (18), 
$|\psi_n(x)|$ diverges to infinity.
when $q=1/2, n=0$, we get $k_0=\pm c=k_*$, so a SDSS at $E=c^2$, for $q>1/2$, we get one or more pairs of CCPEs.
These $k_n$ are complex poles of $t(k)$ or complex zeros of $f(k)=\frac{1}{t(k)}$, in  Fig. 3,as an alternate method, we plot the contours of $\Re(f(k_x+ik_y))=0,$ (red or solid lines) and $\Im(f(k_x+i k_y))=0$ (green or dashed lines) in the plane $(k_x,k_y)$ and note $k_n$ as the point of their inter section for $k_y >0$. In Fig. 3(a) for $q=1/2$, we show un-split SS at $k=\sqrt{2}$ in (b) splitting of SS to a single pair when $q=0.6$ and in (c) splitting SS to five pairs when $q=5$. In all three figs 3, see the single unpaired $k^{\uparrow}$ which is the signature of non-PT-symmetry. Otherwise for the splitting SS in PT-sector [15,16] $k^{\uparrow}=\alpha+i \beta, \alpha, \beta >0$ can be seen to be absent because there the discrete eigenvalues are known [2] to be  either real or complex conjugate pairs.
\begin{widetext} \noindent

%\vspace{-0.99 cm}
\section {\bf one NSDSS, real discrete spectrum and reciprocal reflection}
\begin{figure}
	\includegraphics[width=5 cm,height=5 cm]{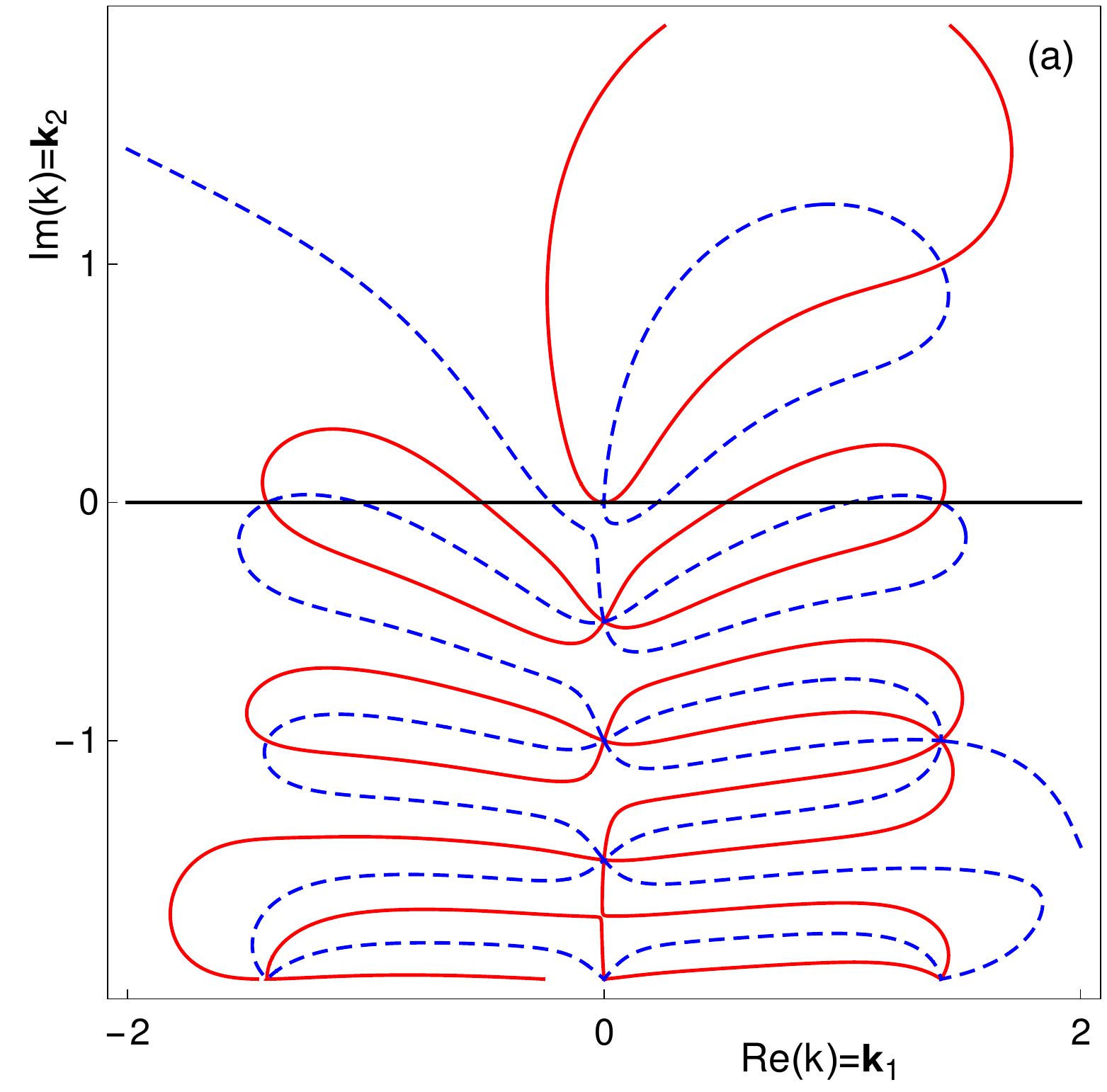}
	\hspace{.1 cm}
	\includegraphics[width=5 cm,height=5 cm]{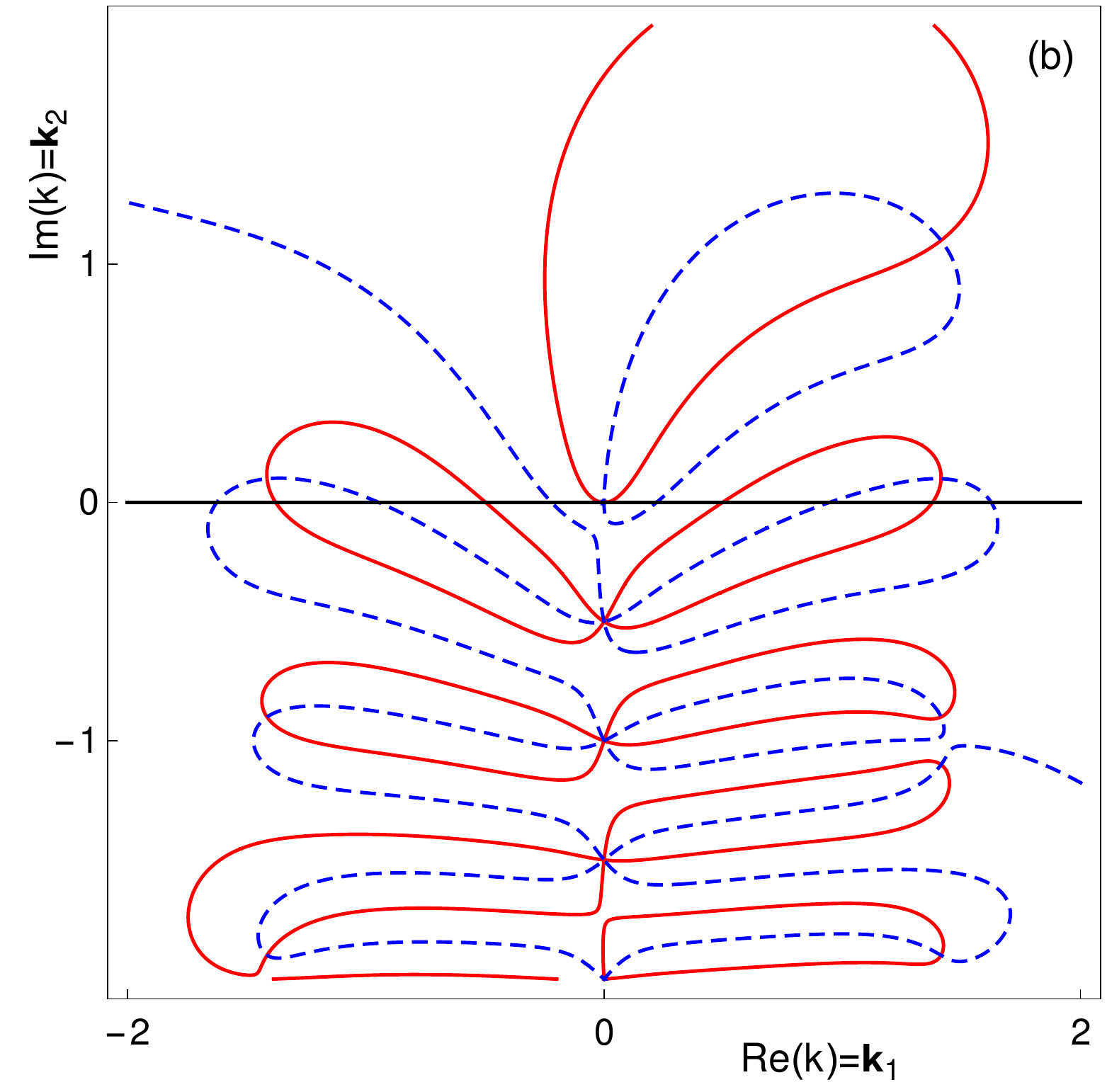}
	\hspace{.1 cm}
	\includegraphics[width=5 cm,height=5 cm]{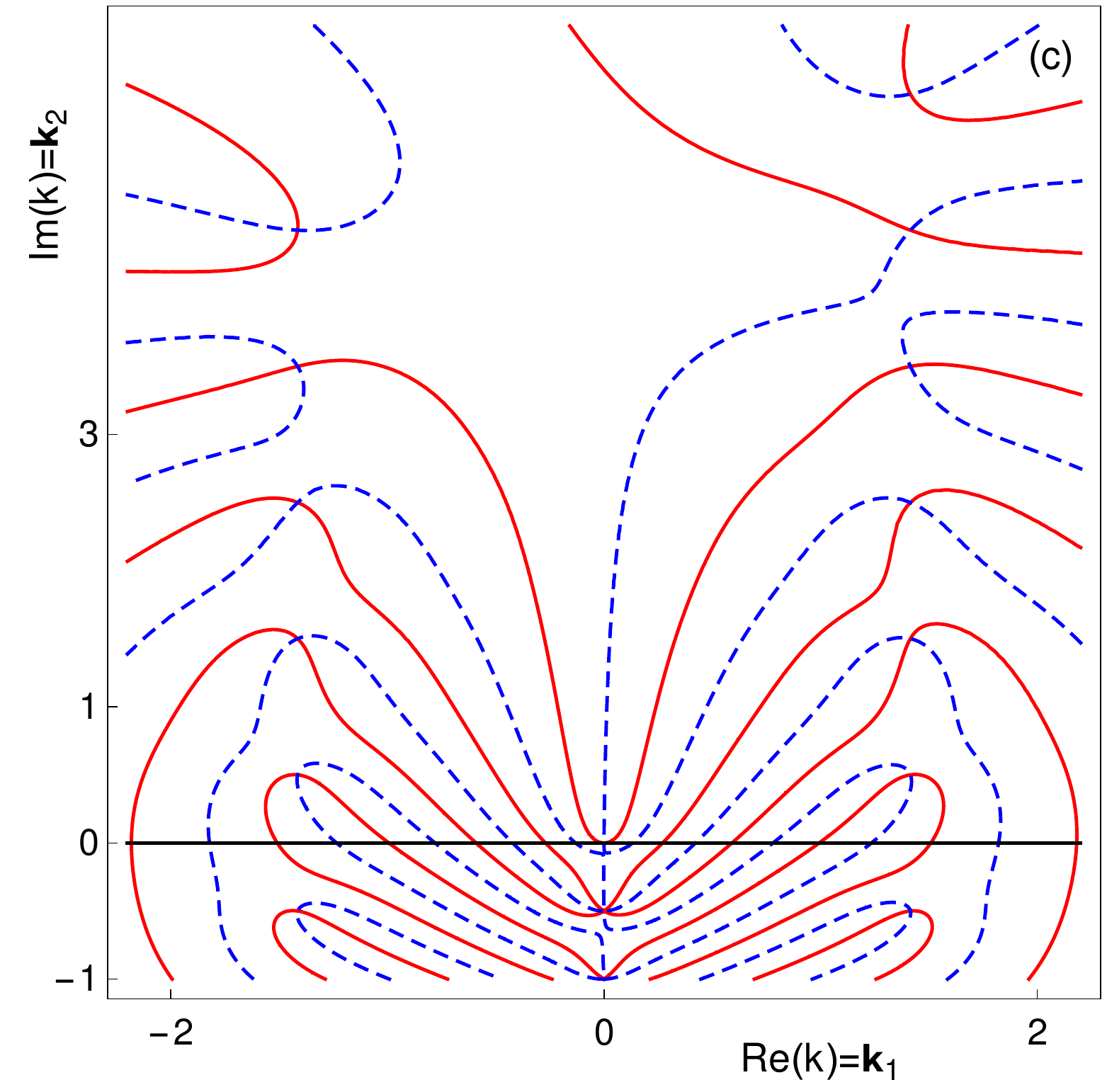}
	\caption{$A=q+1/2-ic, B=c-iq$ splitting of SDSS, the intersection of the solid(red) and dotted (green) curves the giving CCPEs in complex $k$ plane:$k_n=-c+i(q-(n+1/2)]) , 0\le n<q+1/2,(E_n=k_n^2)$. We take $c=\sqrt{2}$, (a): for $q=1/2,  k=k_*=c,$ (b): for $q=0.6$ , notice one pair of CCPE , (c) for $q=5.0$, five pairs of CCPEs. In all three cases see on unpaired complex eigenvalue $k^{\uparrow}=c+(q+1/2)i$ (18).}
\end{figure}
Now, we propose the parametric domain for (1) as ${\bf P_4}:A=-ic$ and $B=id$ where $c,d >0$, the Scarf II potential (1) becomes
\begin{equation}
V(x)=-(d^2+c^2-ic) \mbox{sech}^2 x + (2c d+id) \mbox{sech}{x} \tanh x,
\end{equation}
Which is  non-PT-symmetric. Using this parameterization in $t(k)$% in (3), we get four Gamma functions in the numerator. 
When we set $k=k_{*}=c$, the first Gamma function becomes $\Gamma[0]=\infty$, this gives rise to a single spectral singularity in $t(k)$ or on $T(k)=|t(k)|^2$. This SS is real positive discrete energy embedded in the positive energy continuum. Next, if we set $k=i\kappa_n$ $(\kappa=\sqrt{-E})$ in the argument of the third Gamma function in the numerator of (3) and set
$1/2-d+\kappa_n=-n$, where $n \in I^{+}+\{0\}$, we get the real discrete spectrum ($E<0)$ of (19) as
\begin{equation}
E_n=\kappa^2_n=-(d-1/2-n)^2,\quad n=0,1,2,3,...[d-1/2], \quad \kappa_n=d-1/2+n >0.
\end{equation}
Using the properties of Gamma functions such as $\Gamma(z) \Gamma(-z)=\pi z \mbox{cosec}\pi z$, trigonometric and hyperbolic functions we get the interesting form for the transmission probability as 
\begin{equation}
T(k)=\left|\frac{(k+c)}{(k-c)}~  \frac{\sinh^2 \pi k \cosh^2 \pi k}{(\cosh^2 \pi k -\cosh^2 \pi c)(\sinh^2 \pi k+\cos^2 \pi d)}\right|, \quad E>0,
\end{equation}
clearly showing SS at $E=c^2$ and negative energy bound states at $E_n$ given by Eq. (20).
For the reflection  probabilities, from Eqs. (4-5),  we can write  
\begin{equation}
F_{l,r}(k)=\left[\mp \frac{\sinh \pi c \cos \pi d}{\sinh \pi k} + i \frac{\cosh \pi c \sin \pi d}{\cosh \pi k}\right] \implies |F_l|=F_r|.
\end{equation}
\end{widetext}
Notice that for real values of $c$ and $d$, $F_{l,r}$ cannot become zero for a real value of $k=k_z$, so in this sector we do not get reflectionlessness.
One may also check that the reflection will be reciprocal.
The reflection amplitudes are $r_l(k)=t(k) F_l$ and $r_r(k)=t(k) F_r$  are unequal but their modulus-square being equal, the reflection will be reciprocal ($R_l(k)=R_r(k)$), despite non-Hermiticity and asymmetry of the potential (1). 
The modulus of  the determinant of the $S$-matrix (2) becomes  
\begin{equation}
|\det S(k)|=\left|\frac{k+c}{k-c}\right|  
\end{equation}
At $E=c^2$, $|\det S(c)|=0$ displaying spectral singularity but in time reversed setting $|\det S(-c)|=\infty$ there occurs CPA without lasing [11].
\vspace{-0.7 cm}
\begin{figure}
	\includegraphics[width=4.2 cm,height=4.5 cm]{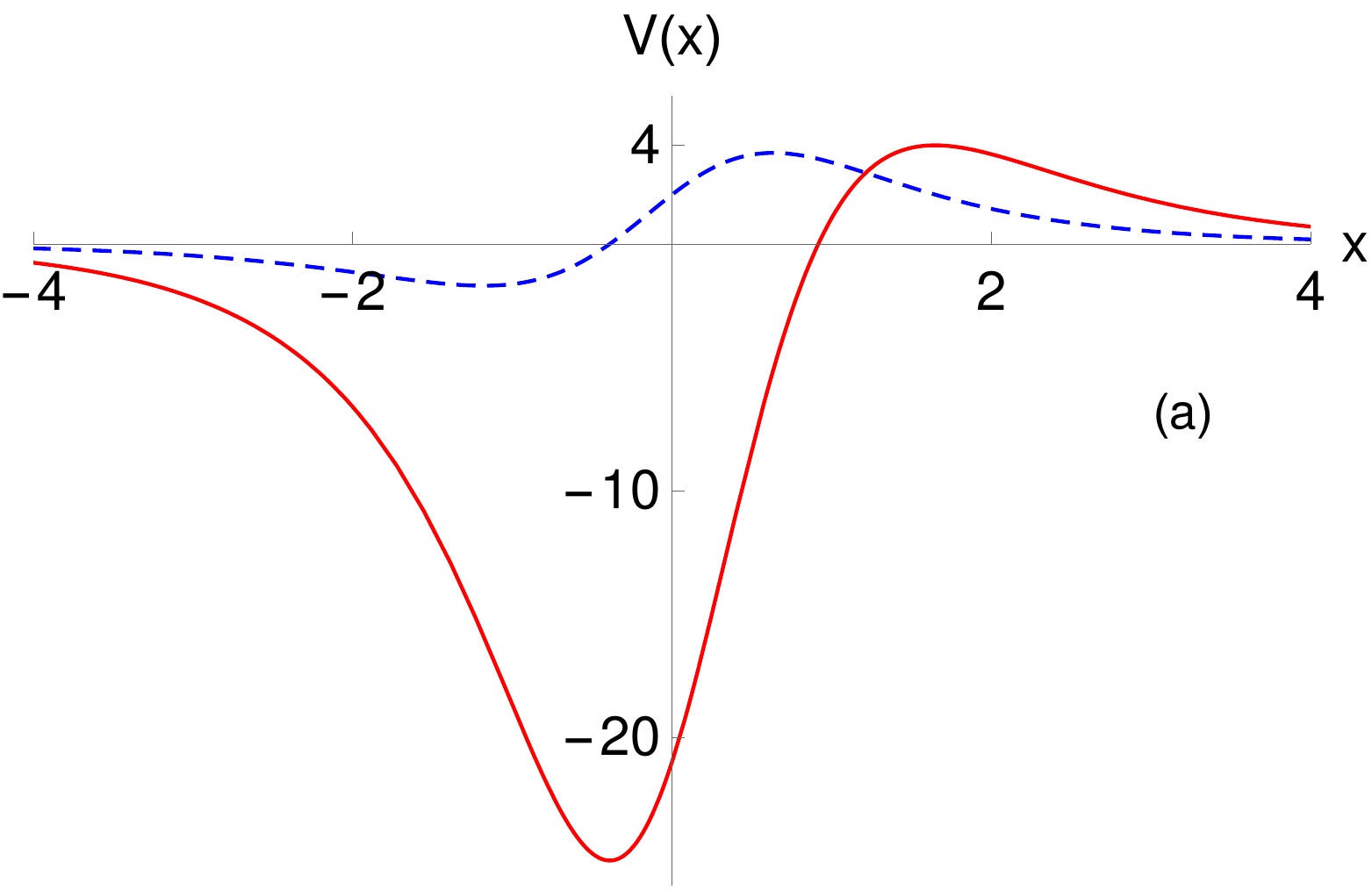}
	\includegraphics[width=4.2 cm,height=4.5 cm]{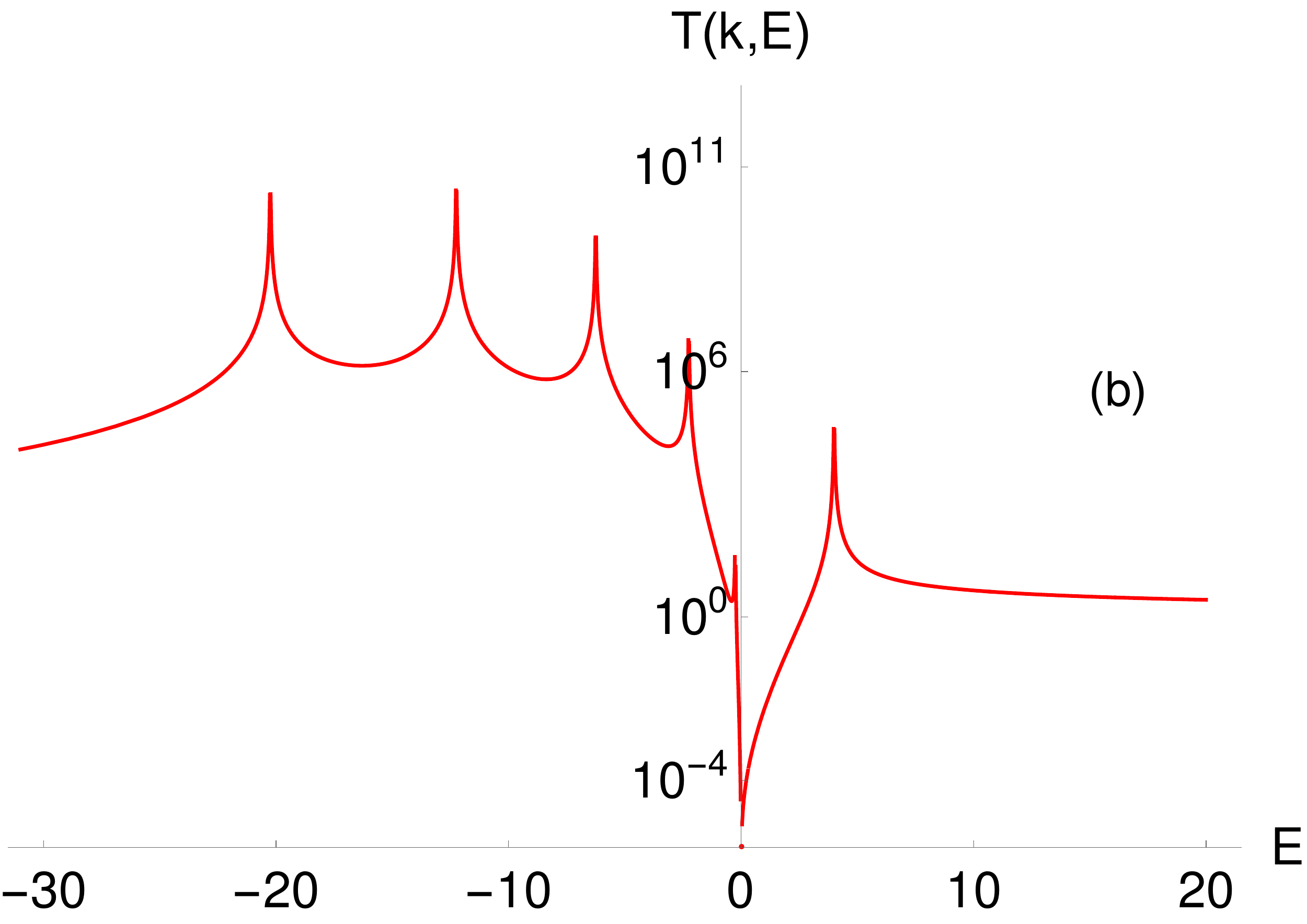}
	\hspace{.1 cm}
	\caption{Depiction of the co-existence of real discrete spectrum and an NSDSS in a fixed non-PT potential when $A =-ic, B=id,  c=2,d =5$. (a): $V(x)$ (b): $T(E<0)(20), T(E>0) (21)$, notice the real  discrete energy poles (see Eq, (20)) at $E_=-\frac{(2n+1)^2}{4}$ and an SS at $E=c^2=4$.}
\end{figure}
\vspace*{0.2 cm}
\section{Discussions}
We find that we can also introduce integers in the parameterization of $A$ and $B$ in (1) at the cost of more involved or un-amenable forms oF $T(k)$. Also in such cases we have not found any other result that differs qualitatively from the results presented here. 
As our search of the non -PT parametric regimes my not have been exhaustive,
so we would like to encourage one to go for new re-parameterizations and use Eqs. (3-5) directly-maybe there arises a new scenario.

Section II, mainly demonstrates the earlier findings [14] analytically, however the new essence is that NSDSS  which gives rise to CPA  exists in non-PT sector and they cannot be split into complex conjugate pairs by perturbing the potential parametrically. On the other hand, the new essence coming out from section III is that a SDSS cannot give rise to CPA but it can be split
such that $E_*(q=0)>Re({\cal E}_n(q))$. Additionally, the sections II and III reveal that that an SS and discrete set of CCPEs cannot co-exist in a non-PT complex potential. This is also true for complex PT-potentials. Unlike the PT-sector, when the SDSS takes place  $T(k_*)=T(-k_*)=\infty$ but $T(k) \ne T(-k) \forall k \in {\cal R}$ (See Fig. 3). It may be recalled that in PT-sector, we have $T(k)=T(-k)$. Our demonstration of the coexistence of real discrete spectrum with an SS in a non-PT potential is unlike PT-symmetric potentials.  The real discrete spectrum is itself remarkable as the complex Scarf II is not known to be pseudo-Hermitian.  

Pseudo-Hermitian (PH) Hamiltonians which are such that  $\eta^{-1} H \eta=H^{\dagger}$[22] may have real discrete or complex conjugate pairs of eigenvalues in two regimes, mutually exclusively. Complex  PT-symmetric potentials  are found to be parity-pseudo-Hermitian. Some non-Hermitian  potentials with imaginary shift of $x$ are PH under $\eta= e^{-p\theta}$, some are so under $\eta=e^{-f(x)}$. Using the idea of PH, very interesting non-Hermitian potentials are constructed [23] which have real discrete or CCPEs. In this regard non-PT Scarf II  may be studied further if it could be  PH under a local metric $\eta(x)$.
\vspace{-0.8cm}
\section{Conclusion:}
Our simple  analytic forms presented in Eqs. (8-11),(13-16), (18), (20-23) are for the non-PT sectors of Scarf II potential which are new and instructive.
We confirm the splitting of self dual spectral singularity in to complex conjugate pairs but unlike the case of PT-sector, here we show a distinct presence of a single (unpaired) discrete complex value  $k^{\uparrow}=\alpha+i \beta, \alpha, \beta >0$ (see Fig. 3).

In non-PT sectors studied here, the new  thought provoking possibilities  which require a confirmation in other complex non-PT potentials are: (1) the existence of one-sided reflectionlessness, (2) self dual spectral singularity not giving rise to coherent perfect absorption (CPA) with or without lasing. (3) non self dual spectral singularity does not split but gives rise to CPA without lasing. (4) a co-existence of a non self dual spectral singularity and real discrete spectrum. In any case here or elsewhere [17,19,21] the existence of real discrete or complex conjugate pairs of eigenvalues hint at replacement of pseudo-Hermiticity with some other more general condition  if Scarf II is not proved to be pseudo-Hermitian.

\vspace{-0.3 cm}
\section*{\large{References}}
%\begin{bibliography}

%\end{bibliography}

\begin{enumerate}
\bibitem{[1]} L. Gendenshtein, {\it JETP Lett.} {\bf 38},  356 (1983)	
\bibitem{[2]} C. M. Bender and S. Boettcher, {\it Phys. Rev. Lett.} {\bf 80},  5243 (1998)
\bibitem{[3]} B. Bagchi and R. Roychowdhury, {\it J. Phys.~A: Math. Gen.} {\bf 33},  L1 (2000)
\bibitem{[4]} Z. Ahmed, {\it Phys. Lett. A} {\bf 282},  343 (2001)
\bibitem{[5]} A. Khare and U. P. Sukhatme, {\it J. Phys. A: Math. Gen.} {\bf 21}, L501  (1988)
\bibitem{[6]} Z. Ahmed,  {\it Phys. Rev. A} {\bf 64},  042716 (2001) 
\bibitem{[7]} Z. Ahmed, {\it Phys. Lett. A} {\bf 324}, 152 (2004)
\bibitem{[8]} A. Mostafazadeh, {\it Phys. Rev. Lett.} {\bf 102},  220402 (2009)
\bibitem{[9]} Z. Ahmed, {\it J. Phys. A: Math. Theor.} {\bf 42}, 472005  (2009)
\bibitem{[10]} B. Bagchi, C. Quesne, {\it J. Phys. A: Math.Gen.} {\bf 43},  395301  (2010)
\bibitem{[11]} Y. D. Chong, L. Ge, H. Cao and A. D. Stone, 
{\it Phys. Rev. Lett.} {\bf 105},  053901 (2010)\\
\bibitem{[12]} S. Longhi, {\it Phys. Rev. A} {\bf 81},  022102  (2010); Y.  Chong Y, Li Ge  and A. D. Stone  {\it Phys. Rev. Lett.} {\bf 106}, 093902 (2011)
\bibitem{[13} A. Mostafazadeh, {\it J. Phys. A: Math. Theor.} {\bf 45},  444024 (2012)
\bibitem{[14]} Z. Ahmed, {\it J. Phys. A: Math. Theor.} {\bf 45}, 032004  (2012)
\bibitem{[15]} V. V. Konotop and D. A. Zezyulin, {\it Optics Letters} {\bf 42}, 5206 (2017)
\bibitem{[16]} Z. Ahmed, S. Kumar and D. Ghosh, {\it Phys. Rev. A} {\bf 98}, 042101 (2018)
\bibitem{[17]}  D. A. Zezyulin and   V. V. Konotop, {\it New J. Phys.} {\bf 22}, 013057 (2020) 
\bibitem{[18]} V. V. Konotop, E. Lashtanov, and B. Vainberg, {\it Phys. Rev. A} {\bf 99}, 043838 (2019) 
\bibitem{[19]} A. Mostafazadeh, {\it Phys. Rev. A} {\bf 87}, 012103 (2013); B. Midya, {\it Phys. Rev. A} {\bf 89}, 032116 (2014)
\bibitem{[20]} Z. Ahmed, {\it J. Phys. A: Math. Theor.} {\bf 42}, 485303 (2014).
\bibitem{[21]} Z. Ahmed and J. A. Nathan, {\it Phys. Lett. A} {\bf 379}, 865 (2015)
\bibitem{[22]} A. Mostafazadeh, {\it J. Math. Phys.}  {\bf 43}, 286, {\bf 43} 2814 (2002); Z. Ahmed, {\it Phys. Lett. A} {\bf 290}, 19 (2001); {\bf 294}, 287 (2002).
\bibitem{[23]} T.V. Fityo, {\it J. Phys. A: Math. Theor.} {\bf 35}, 5813 (2002);
C. Hang, G. Gabadadze, and  G. Huang, {\it Phys. Rev. A} {\bf 95}, 023833 (2017)




\end{enumerate}
\end{document}